\pdfoutput=1

\documentclass{iopart}

\usepackage{epsfig}
\usepackage{amscd,amssymb}
\usepackage{graphicx,xspace}
\usepackage{pstricks}
\usepackage[left=3cm,top=3cm,right=3cm,bottom=3cm]{geometry}
\usepackage{color}

\begin{document}

\title{Winding of planar gaussian processes}

\author{Pierre Le Doussal}
\address{CNRS-Laboratoire de Physique Th\'eorique de l'Ecole Normale Sup\'erieure, 24 rue Lhomond, 75231 Paris Cedex-France}

\author{Yoav Etzioni and Baruch Horovitz }
\address{Department of Physics, Ben Gurion University, Beer Sheva 84105 Israel}

\begin{abstract}
We consider a smooth, rotationally invariant, centered gaussian process in the plane, with arbitrary correlation matrix $C_{t t'}$. We study the winding angle $\phi_t$ around its center. We obtain a closed formula for the variance of the winding angle as a function of the matrix $C_{tt'}$. For most stationary processes $C_{tt'}=C(t-t')$ the winding angle exhibits diffusion at large time with diffusion coefficient $D =  \int_0^\infty ds C'(s)^2/(C(0)^2-C(s)^2)$. Correlations of $\exp(i n \phi_t)$ with integer $n$, the distribution of the angular velocity $\dot \phi_t$, and the variance of the algebraic area are also obtained. For smooth processes with stationary increments (random walks) the variance of the winding angle grows as $\frac{1}{2} (\ln t)^2$, with proper generalizations to the various classes of fractional Brownian motion. These results are tested numerically. Non integer $n$ is studied numerically.
\end{abstract}

\maketitle

\section{Introduction and model}

The winding of planar random processes has been studied for a while. These are of interest for the physics of polymers \cite{polymers,grosberg}, flux lines in superconductors \cite{super,kardar} and quantum Hall effect \cite{QHE}. Recently there was revived interest in winding properties of processes described by Schramm-Loewner Evolutions (SLE) \cite{sle}, such as the loop erased random walk \cite{hagendorf}. In the case of the planar Brownian motion the distribution of the winding angle $\phi_t$ around a point $O$ was computed a long time ago by Spitzer \cite{spitzer} who found that $\phi_t/\ln t$ has a Cauchy distribution, i.e with infinite first moment. This peculiar feature was later understood to be related to the large winding accumulated while the trajectory wanders infinitely close to point $O$, and is removed by considering either a small excluded region around $O$, or a lattice cutoff, or some other regularization of the short time, and leading instead to exponentially decaying distributions \cite{bmwinding,duplantier,next,comtet}. Recently it was found that correlated gaussian processes related to the fractional Brownian motion, a scale invariant process with stationary increments correlated in time, has similar properties \cite{rajabpour}.

The aim of this paper is to study the winding of a very general continuous-time gaussian process $\xi_t=\xi^x_{t} + i \xi^y_{t}$ in the complex plane with arbitrary correlations in time. The only restriction, mainly to avoid cumbersome formula, is that the measure is rotationally invariant around the origin $0$ and the winding angle $\phi_t$ is measured around point $0$, i.e. $\xi_t=r_t e^{i \phi_t}$ where $\phi_t$ is a continuous real function of time and $r_t=|\xi_t|$. The process is thus centered $\langle \xi_t \rangle = 0$ and fully characterized by its two-time correlation function:
\begin{eqnarray}
\langle \xi^i_t \xi^j_{t'} \rangle = \delta_{ij} C_{tt'}  \label{defmodel}
\end{eqnarray}
with $i,j=x,y$, equivalently $\langle \xi_t \xi_{t'} \rangle=0$ and $\langle \xi_t^* \xi_{t'}\rangle=2 C_{tt'}$. The most general form would be $\langle \xi_t^* \xi_{t'} \rangle=2 (C_{tt'} + i A_{tt'})$ but we also assume reflection symmetry which forbids the antisymmetric term $\epsilon_{ij} A_{tt'}$ in (\ref{defmodel}) with $\epsilon_{12}=-\epsilon_{21}=1$. Since the process $\xi_t/\sqrt{C_{tt}}$ has the same winding angle as $\xi_t$, observables involving only the winding angle should only depend on the combination:
\begin{eqnarray}
c_{tt'} = C_{tt'}/\sqrt{C_{tt} C_{t't'}}
\end{eqnarray}
with $c_{tt}=1$ and from Cauchy-Schwartz inequalities, $|c_{tt'}| \leq 1$, the bound being saturated, i.e. $c_{tt'}=\pm 1$, if and only if $\xi_t=\pm \sqrt{\frac{C_{tt}}{C_{t't'}}} \xi_{t'}$. Some particular cases are (i) stationary process $C_{tt'}=C(t-t')$ and one defines $c_{t0}=c(t)=C(t)/C(0)$ (ii) process with stationary increments $C^{(1,1)}_{tt'}:=\partial_{t}\partial_{t'} C_{tt'}=C_2(t-t')$ (here and below we adopt the following definition for partial derivatives $\partial_t C_{tt'}=C^{(1,0)}_{tt'}$ etc..).  (Strict) normalizability of the Gaussian measure requires that these functions have (strictly) positive Fourier transforms $\tilde c(\omega) \geq 0$ and $\tilde C_2(\omega) \geq 0$. In some cases we restrict to a process which everywhere below we call "smooth", meaning - by definition here - $\xi_t$ differentiable at least once, i.e. $C^{(1,1)}_{tt}$ exists (equal $- C''(0)$ for a stationary process). For such a smooth process, $\langle \dot \xi^i_t \xi^i_t \rangle= C^{(1,0)}_{tt}=C^{(0,1)}_{tt}=\frac{1}{2} \partial_t C_{tt}$, which vanishes if furthermore the process is stationary. These processes appear everywhere in physics, prominent examples occur in the study of quantum noise, where $\tilde C(\omega)=\eta \hbar \omega \coth(\hbar \omega/2 T)$ (supplemented by a bath-dependent short time cutoff, e.g. at $T=0$, $\tilde C(\omega)=\hbar \eta |\omega| e^{-\tau_0 |\omega|}$, i.e
 $C(\tau)=\frac{\eta \hbar}{\pi} \frac{\tau_0^2-\tau^2}{(\tau_0^2+\tau^2)^2}$) or of quantum Brownian motion \cite{quantumB}
and has served as a motivation for the present work \cite{us}

The outline of the paper is as follows. In Section 2 we study single time quantities. The distribution of angular velocity is obtained. In Section 3 we study the periodized winding probability distribution which is easier than the full one. The correlations of $\exp(i n \phi_t)$ are obtained analytically for integer $n$, and studied numerically also for non-integer $n$. In Section 4 we obtain a closed formula for the variance of the winding angle as a function of the matrix $C_{tt'}$. We show that 
for most stationary processes the winding angle exhibits diffusion at large time and we obtain the diffusion coefficient. We also study non-stationary process such as the random walk and the various classes of fractional Brownian motion. Finally in Section 5 the variance of the algebraic area is obtained. Most results are tested numerically. 

\section{Single time quantities}

Single time quantities are easily extracted from the Gaussian distribution $\sim d^2 \xi_t e^{-|\xi_t|^2/(2 C_{tt})}$ performing change of variables. Everywhere below $d^2 \xi=d\xi d\xi^*=d\xi^x d\xi^y$. The modulus is distributed as $P(r_t) dr_t$ with $P(r) = \frac{r}{C_{tt} } e^{-r^2/(2 C_{tt})}$, hence the probability to be within $r_t< \epsilon$ near the center vanishes as $\epsilon^2/(2 C_{tt})$. To compute the distribution of the angular velocity $\dot \phi_t$ one uses that $X_t=(\xi_t,\dot \xi_t)$
is gaussian with measure  $\frac{d^2 \xi_t d^2 \dot \xi_t}{(2 \pi)^2} \det(M)  e^{- \frac{1}{2} X^* M X}$ and correlation matrix $M^{-1}=((C_{tt},C^{(1,0)}_{tt}),(C^{(0,1)}_{tt},C^{(1,1)}_{tt}))$. Let us denote $\dot \xi_t = \alpha_t \xi_t$ with $\alpha_t=\dot r_t/r_t + i \dot \phi_t$. Here we have requested a smooth process.The measure becomes $\frac{d^2 \xi_t d^2 \alpha_t}{(2 \pi)^2} |\xi_t|^2 \det(M)  e^{-  \frac{1}{2} \beta |\xi_t|^2 }$ 
where $\beta=(1,\alpha_t^*) M (1,\alpha_t)$. Integration over $\xi_t$ yields
the joint distribution $P(\dot \rho_t,\dot \phi_t) d\dot \rho_t d \dot \phi_t$, with $\rho_t=\ln r_t$, equal to:
\begin{eqnarray}
&& \frac{d\dot \rho_t d \dot \phi_t}{\pi} \frac{C_{tt} C^{(1,1)}_{tt}}{(C^{(1,1)}_{tt} - 2 C^{(1,0)}_{tt} \dot \rho_t + C_{tt} ( \dot \rho_t^2 + \dot \phi_t^2))^2}
\end{eqnarray}
Integration yields:
\begin{eqnarray}  \label{velocitydistrib}
&& P(\dot \phi_t) d \dot \phi_t = d \dot \phi_t  \frac{a_t}{2 (a_t + \dot \phi_t^2)^{3/2}} 
\end{eqnarray}
with $a_t=(C_{tt} C^{(1,1)}_{tt} - (C^{(1,0)}_{tt})^2)/C_{tt}^2 = \partial_t \partial_{t'} \ln |c_{tt'}||_{t'=t}$. For a stationary process $a_t=a=-c''(0)$. For stationary increments
$a_t = C_2(0)/C_{tt} - \frac{1}{4} (\partial_t \ln C_{tt})^2$. Note that this distribution is broad, it does have a first moment but no second moment i.e. $\langle \dot \phi_t^2 \rangle$ is infinite. 

\section{Periodized winding}

Next one can compute two time correlations of the winding angle. The two time probability measure of the process 
can be written:
\begin{equation}
 \frac{r_t r_{t'} dr_t dr_{t'} d\phi_t d\phi_{t'} }{(2 \pi)^2 \Delta_{tt'}}  \exp(-\frac{C_{t't'} r_t^2 + C_{tt} r_{t'}^2 - 2 C_{tt'} r_t r_{t'} \cos(\phi_t-\phi_{t'})}{2 \Delta_{tt'}} )
\end{equation}
with $\Delta_{tt'}=C_{tt} C_{t't'} - C_{tt'}^2$ 
hence integration over $r_t$ and $r_{t'}$ allows to obtain the probability distribution of $\cos(\phi_t-\phi_{t'})$. Equivalently this gives the probability of $\phi := \phi_t-\phi_{t'}$ modulo $2 \pi$, i.e it gives the periodized probability $\tilde P(\phi)=\sum_{m=-\infty}^{+\infty} P(\phi+2 \pi m)$ where $P(\phi)$ is the probability of the total winding $\phi \in ]-\infty,+\infty[$. Defining:
\begin{equation}
 F(z) := \int_{x,y>0} x y e^{-\frac{x^2}{2} - \frac{y^2}{2} + x y z}  = \frac{1}{1-z^2} +  \frac{z \arccos (-z)}{(1-z^2)^{3/2}}  
\end{equation}
for $-1< z <1$, with $\arccos (-z) = \pi - \arccos z$ and $F(z)= \sum_{n=0}^\infty \frac{2^n}{n!} \Gamma[1+ \frac{n}{2}]^2 z^n =  1 + \frac{\pi z}{2} + 2 z^2 + O(z^3)$, 
one finds:
\begin{eqnarray} \label{distribcos} 
&& \tilde P(\phi) = \frac{1}{2 \pi} (1- c_{tt'}^2)  F( c_{tt'} \cos(\phi) ) 
\end{eqnarray}
One can check that $\int_0^{2 \pi} d\phi \tilde P(\phi)=1$. An interesting limit is $t$ close to $t'$. Then $c_{tt'}$ is close to unity and using the expansion 
$F(z) = \frac{\pi}{2 \sqrt{2} (1-z)^{3/2}} - \frac{\pi}{8 \sqrt{2 (1-z)} } + O(1)$ we find that:
\begin{eqnarray} \label{distrib2}
&&  \tilde P(\phi) \approx \frac{1-c_{tt'}}{(2(1-c_{tt'}) + \phi^2)^{3/2}}
\end{eqnarray}
and furthermore we expect that $P(\phi) \approx \tilde P(\phi)$ since the probability of $2 \pi n$ winding with $n \neq 0$ is negligible in that limit. Upon Taylor expansion in $t-t'$ one finds that this result is consistent with (\ref{velocitydistrib}) but the approach here is more general as the formula (\ref{distribcos}) does not require a smooth process. The only assumption in (\ref{distrib2}) is then the continuity of $c_{tt'}$. The short time behaviour is controled by the distribution (\ref{distrib2}) for moments with $0<\alpha<2$, i.e. $\langle |\phi|^\alpha \rangle_{\tilde P} \approx K_\alpha (1-c_{tt'})^{\alpha/2}$ with $K_\alpha=2^{\alpha/2} \Gamma(1-\frac{\alpha}{2}) \Gamma(\frac{1+\alpha}{2})/\sqrt{\pi}$, and becomes dominated by the cutoff at $\phi=O(1)$ for $\alpha \geq 2$ with 
$\langle |\phi|^\alpha \rangle_{\tilde P} \approx K'_\alpha (1-c_{tt'})^{2-\frac{\alpha}{2}}$. For instance the variance of the winding angle is found as 
\begin{equation}
\langle \phi^2 \rangle_{\tilde P} \approx - (1-c_{tt'}) \ln(1-c_{tt'})
\end{equation}
for $c_{tt'}$ close to unity,
and we check below that this coincides with the behavior of $\langle \phi^2 \rangle_{P}$ at short time differences. 

This allows to compute the correlation functions ${\cal C}_n(t,t')
= \langle e^{ i n (\phi_t - \phi_{t'}) } \rangle$ for {\it integer} $n$, which thus have closed expressions as a function of the matrix $C_{tt'}$:
\begin{eqnarray} \label{final}
&& {\cal C}_n(t,t') = F_n(c_{tt'}) 
\end{eqnarray}
One finds for instance:
\begin{eqnarray}
&& F_1(c)=\frac{1}{c} (E(c^2) + (c^2-1) K(c^2)) \\
&& F_2(c) = 1 + (\frac{1}{c^2}-1) \ln(1-c^2) 
\end{eqnarray}
and we obtain the limiting behaviours:
\begin{eqnarray}
&&  F_1(c)= \frac{\pi c}{4} + \frac{\pi c^3}{32} + \frac{3 \pi c^5}{256}+O(c^7) \\
&& = 1 - \frac{1-c}{2} (\ln(\frac{8}{1-c}) -1) + O(\ln(1-c) (1-c)^2) \nonumber
\end{eqnarray}
for small $c$ (large time separation), and for $c$ near 1 (small time separation), respectively,
and $F_2(c)=\frac{c^2}{2} + O(c^4)$ and more generally $F_n(c) = \frac{\Gamma[1+ \frac{n}{2}]^2}{n!} c^n + O(c^{n+2})$ for integer $n$ at small $c$. 
 
\begin{figure}[htpb]
  \centering
  \includegraphics[width=0.6\textwidth]{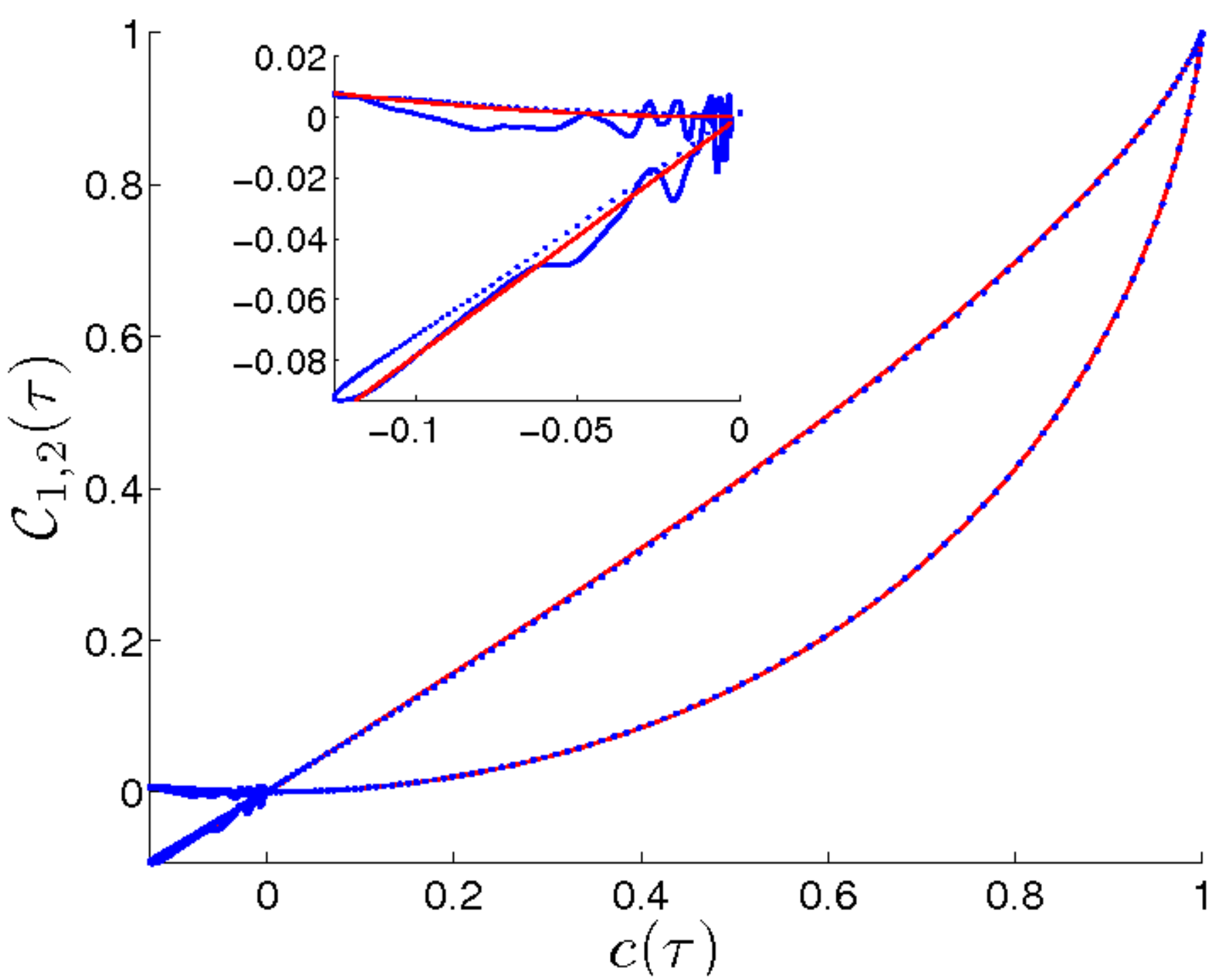}

  \caption{The correlation function $\mathcal{C}_n(\tau)$ for $n=1$ (top curve) and for $n=2$ (bottom curve) as a function of $c(\tau)$ for $c(\tau)=\frac{1-\tau^2}{(1+\tau^2)^2}$ (Blue curves) and the prediction for $F_1(c)$ and $F_2(c)$ (Red curves). The inset shows how the curve is retraced for negative values of $c$.}
    \label{fig:1}
\end{figure}

We have checked these results numerically for several stationary processes where $c_{tt'}=c(\tau)=C(\tau)/C(0)$ where $\tau=t-t'$.
The process $\xi_t^i$ was generated numerically using a discrete Fourier transform of $\sqrt{\tilde{c}(\omega)N\Delta\tau} \mathcal{A}^i$, where $N$ the number of points is typically $N=2^{16}$, $\Delta \tau=.01$ is the time segment in the process and $\mathcal{A}^i$ is a unit white gaussian process. We have computed ${\cal C}_n(\tau)$ where the average $\langle e^{ i n \phi } \rangle$ is over the time range and over several realizations, typically 10. We have plotted ${\cal C}_n(\tau)$ parametrically as a function of $c(\tau)$ for various type of noises. Up to numerical accuracy all the curves fall on the predicted master curve ${\cal C}_n(\tau)=F_n(c(\tau))$. When $c(\tau)$ is non monotonous, the master curve may be traced more than once. This is illustrated in Fig. \ref{fig:1}.

\section{Variance of the total winding angle}

\begin{figure}[htpb]
  \centering
  \includegraphics[width=0.6\textwidth]{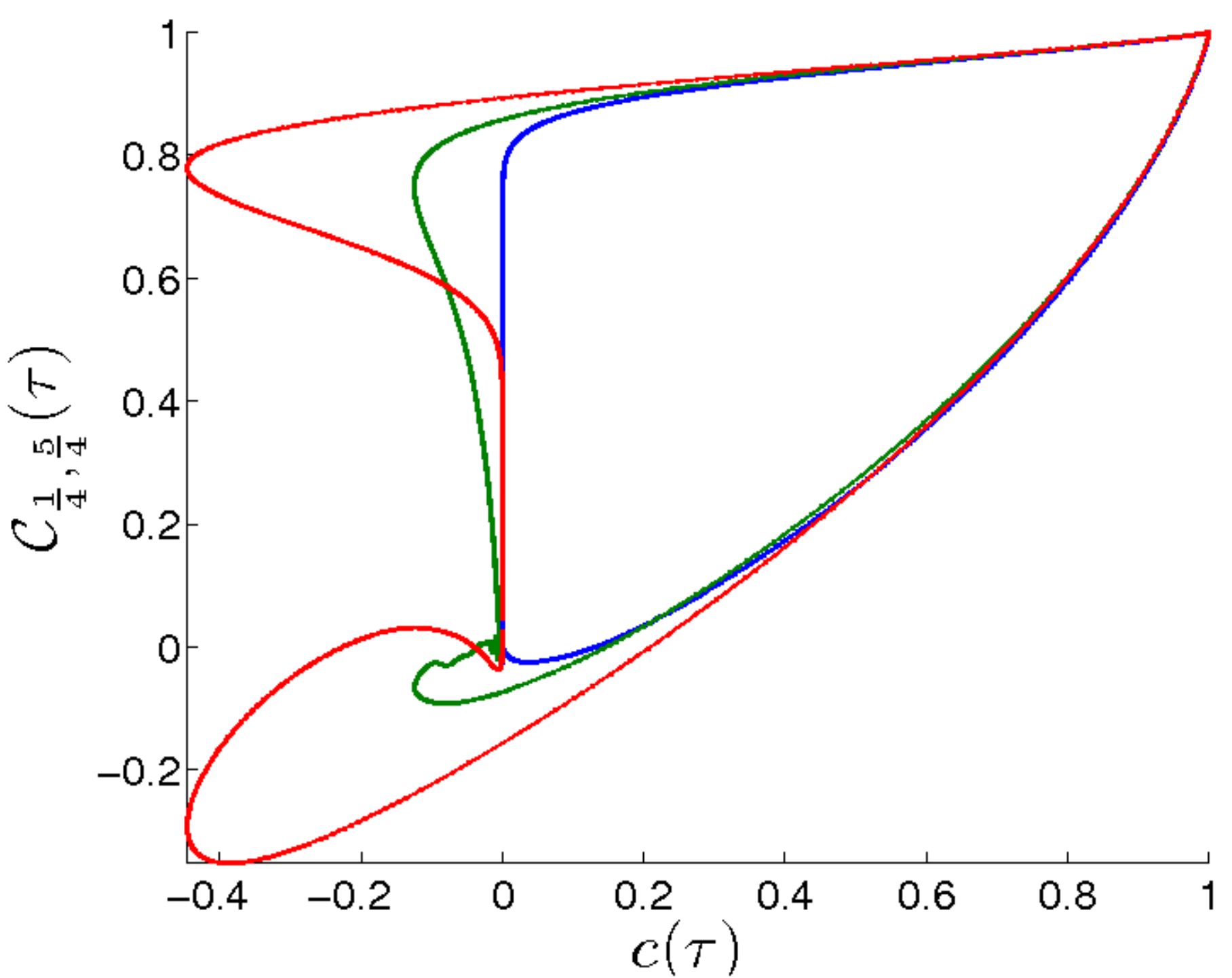}

  \caption{The correlation function $\mathcal{C}_n(\tau)$ for $n=\frac14$ (three curves starting on the top) $\frac54$ (three curves starting on the bottom) as a function of $c(\tau)$ for three processes: $c(\tau) = \exp{-\tau^2/2}$ (in blue), the $c(\tau)$ used in Fig.1 (in green), $c(\tau) = (1-\tau^2)\exp{-\tau^2/2}$ (in red). Note that for each $n$ the three curve remain very close for $c>0.4$ and that for $n=5/4$ all processes change sign}
    \label{fig:2}
\end{figure}

The previous results are easy to derive, and are simple functions of $c_{tt'}$, but they do not contain information about integer winding. They only probe $\tilde P(\phi)$, the periodized winding angle distribution. An interesting question is how to access the full winding distribution $P(\phi)$ and whether its dependence on the matrix $C_{tt'}$ remains tractable. It is a more difficult question since to compute the full winding angle one must follow somehow the time evolution of the process, e.g. use that $\phi = \phi_{t}-\phi_{t'}=\int_{t'}^t d\phi_s$. A related difficult question, which requires the full distribution $P(\phi)$, is to obtain the averages ${\cal C}_n(t,t')= \langle e^{ i n (\phi_t - \phi_{t'}) } \rangle$ for {\it non integer} $n$. It is seen on Fig. \ref{fig:2} that these are not simple functions, but rather unknown and 
more complicated functionals, of $c_{tt'}$.

Here we present the simplest result on this question, the variance of the winding angle. Here, for simplicity, and to avoid stochastic calculus subtelties, we first restrict to a smooth, i.e. differentiable process as discussed above. We only need to compute the two time angular velocity correlation ${\cal C}_v(t,t')=\langle \dot \phi_{t} \dot \phi_{t'} \rangle$. We use that $\dot \phi_t = {\rm Im}(\dot \xi \xi^*)/|\xi|^2=
\epsilon_{ij} \xi^i_t \dot \xi^j_t/|\xi_t|^2$ hence:
\begin{eqnarray}
&& {\cal C}_v(t,t') = 2   \int_{s,s'>0} G_{ss'tt'} \\
&& G_{ss'tt'} = \langle \dot \xi^y_t \dot \xi^y_{t'} \rangle_{s,s'} \langle  \xi^x_t \xi^x_{t'} \rangle_{s,s'} -  \langle \dot \xi^y_t  \xi^y_{t'} \rangle_{s,s'} \langle  \xi^x_t \dot \xi^x_{t'} \rangle_{s,s'} \nonumber
\end{eqnarray}
using isotropy, where $\langle O[\xi_x] \rangle_{s,s'}= \langle O[\xi_x] e^{-s (\xi^x_t)^2 - s' (\xi^x_{t'})^2} \rangle$ and the same for averages over $\xi_y$. The integrals over $s$ restore the (difficult) denominators. Each average can be computed from the generating function:
\begin{eqnarray}
&& \langle e^{i \mu \xi^x_t + i \mu' \xi^x_{t'} + i \mu_1 \xi^x_{t_1} + i \mu_2 \xi^x_{t_2}} \rangle = e^{- \frac{1}{2} (C_{tt} \mu^2 + C_{t't'} {\mu'}^2) - \mu \mu' C_{tt'}} \nonumber \\
&& \times e^{- \mu_1 \mu_2 C_{t_1 t_2} - \sum_{i=1,2}  \frac{1}{2} C_{t_i t_i} \mu_i^2 + \mu_i (\mu C_{t_it} + \mu' C_{t_i t'} ) }
\end{eqnarray}
by taking appropriate derivatives w.r.t. $\mu_i$ and $t_i$ at $\mu_i=0$ and coinciding times, and integrating with the measure 
 $\frac{1}{4 \pi \sqrt{s s'}} \int_{\mu,\mu'} 
e^{-\frac{\mu^2}{4 s}-\frac{(\mu')^2}{4 s'}}$ to restore the $\langle ... \rangle_{s,s'}$ averages. After a straighforward but lengthy algebra, and massive simplifications, one finds:
\begin{eqnarray}
 && G_{ss'tt'} = \frac{- C^{(1,0)}_{tt'} C^{(0,1)}_{tt'} + C^{(1,1)}_{t t'} C_{t,t'} }{
(1 + 2 s C_{tt} + 2 s' C_{t't'} + 4 s s' (C_{tt} C_{t't'} - C_{t t'}^2))^{2}} \nonumber 
 \end{eqnarray} 
which, after integration over $s,s'$ gives the angular velocity correlation, which is our main result:
\begin{eqnarray}
&&  {\cal C}_v(t,t') : =\langle \dot \phi_{t} \dot \phi_{t'} \rangle = \frac{1}{2} (\frac{- C^{(1,1)}_{tt'} C_{tt'} + C^{(1,0)}_{tt'} C^{(0,1)}_{tt'} }{C_{tt'}^2} )  \ln(1 - c_{tt'}^2) \nonumber  \\
&& = - \frac{1}{2} (\partial_t \partial_{t'} \ln |c_{tt'}|)  \ln(1 - c_{tt'}^2) \label{result}
\end{eqnarray} 
where we recall that $c_{tt'}=C_{tt'}/\sqrt{C_{tt}C_{t't'}}$. The variance of the winding angle is then obtained as:
\begin{eqnarray}
\Phi_{tt'} = \langle (\phi_t-\phi_{t'})^2 \rangle = \int_{t'}^t dt_1 \int_{t'}^t dt_2  {\cal C}_v(t_1,t_2)  \label{integrate}
 \end{eqnarray} 
hence $\partial_t \Phi_{tt'}=2 \int_{t'}^t dt_2  {\cal C}_v(t,t_2)$, where ${\cal C}_v(t_1,t_2)$ is given by (\ref{result}). We now discuss separately stationary and non-stationary processes. 

\subsection{stationary processes} 

Let study first stationary processes $c_{tt'}=c(\tau)=C(\tau)/C(0)$ with $\tau=t-t'$. Then the angular velocity correlation becomes ${\cal C}_v(t,t') = {\cal C}_v(t-t')$ with:
\begin{eqnarray} \label{eqcv} 
&&  {\cal C}_v(\tau) =  \frac{1}{2} (\frac{C''(\tau) C(\tau) - C'(\tau)^2}{C(\tau)^2} )  \ln(1 - c(\tau)^2) \nonumber  \\
&& =  \frac{1}{2} (\partial_\tau^2 \ln |c(\tau)|)  \ln(1 - c(\tau)^2)
\end{eqnarray} 
which exhibit a divergence at small time $\tau$, ${\cal C}_v(\tau) \approx c''(0) \ln(\tau \sqrt{-c''(0)})$, but an integrable one.
The winding angle variance takes the form $\Phi_{tt'} = \Phi(t-t')$, and using that $\partial_\tau \Phi(\tau)=2 \int_{0}^\tau d\tau_2  {\cal C}_v(\tau_2)$ one finds upon integration by part:
\begin{equation} \label{derivative2}
\partial_\tau \Phi(\tau)= 2 \int_0^\tau ds \frac{c'(s)^2}{1-c(s)^2} + \frac{c'(\tau)}{c(\tau)} \ln(1-c(\tau)^2)
\end{equation}
with no boundary term at $\tau=0$ since $c(\tau)\approx1+\frac{1}{2} c''(0) \tau^2$ at small $\tau$, with $c''(0)<0$ since the process is smooth. 
The small $\tau$ behaviour of the variance of the winding angle is $ \Phi(\tau) \approx \tau^2 (c''(0) \ln(\tau \sqrt{-c''(0)}) - \frac{3}{2} c''(0))$. More generally, the formula (\ref{derivative2}) holds for processes such as $c(\tau)=e^{-\tau^a}$ at small $\tau$ with $a>1$ so that the small time singularity of ${\cal C}_v(\tau)$ be integrable. 

If we now consider processes such that $c(+\infty)=0$ then we find that the generic behavior is that the winding angle {\it diffuses} at large time as $\Phi(\tau) \sim 2 D \tau$ with a diffusion coefficient:
\begin{eqnarray}  \label{diffusion}
&& D =  \int_0^\infty ds \frac{c'(s)^2}{1-c(s)^2} 
\end{eqnarray}
an integral which converges at small $s$ when the process is smooth since then $c'(0)=0$. In fact, this formula, as well as (\ref{derivative2}) holds also for some processes with $c''(0)=+\infty$, e.g. such as $c(\tau)=e^{-\tau^a}$ at small $\tau$ with $a>1$, the main condition being that the small time singularity of ${\cal C}_v(\tau)$ be integrable. The convergence at large $s$ should be discussed separately. Since $D>\int ds c'(s)^2$ a necessary condition for convergence at large $s$ is $\int ds c'(s)^2 = \int \frac{d\omega}{2 \pi} \omega^2 \tilde c(\omega)^2  < + \infty$. For a positive $c(s)$ this is guaranteed by $c(\infty)=0$. For oscillating $c(s)$, e.g. $c(s)=cos(s) f(s)$ it requires $|f(s)|$ to decay faster than $1/\sqrt{s}$, or in Fourier, e.g. if $c(\omega) \sim 1/|\omega-\omega_c|^b$ then one must have $b<1$. Interestingly, (\ref{diffusion}) can also be written as:
\begin{eqnarray}  \label{diffusion2}
&& D =  \int_0^\infty ds (\frac{d\theta}{ds})^2
\end{eqnarray}
where $c(s)=\sin \theta(s)$ with $\theta(s) \in [-\pi/2,\pi/2]$, $\theta(0)=\pi/2$ and here $\theta(+\infty)=0$.
From there one sees that the strict criterion for convergence is that $\int \frac{d\omega}{2 \pi} \omega^2 c_\theta(\omega)^2 < + \infty$ where 
$\tilde c_\theta(\omega)$ is the Fourier transform of $c_\theta(\tau)=\langle \theta(0) \theta(\tau) \rangle$, and is also positive (the variable $\theta$ however is not Gaussian).

Let us give examples of some of the non-generic situations where winding angle diffusion does not occur. The simplest case is the process $\xi_t = \xi_1 e^{i t} + \xi_2 e^{-i t}$ with $\xi_1$ and $\xi_2$ two i.i.d. complex gaussian noises, which has correlation $c_{tt'}=\cos(t-t')$. One finds $C_v(\tau)=- \frac{ \ln \sin^2 \tau}{2 \cos^2 \tau}$ and $\partial_\tau \Phi(\tau)=2 \tau - \tan \tau \ln(\sin^2 \tau)$, hence the winding angle grows faster than diffusive as $\Phi(\tau) \sim \tau^2$. Consider next $c(\tau)=J_0(\tau)$, i.e. $c(\omega) \sim \theta(1-\omega^2) (1-\omega^2)^{1/2}$. The integral (\ref{diffusion}) is log-divergent at large $s$ and one finds superdiffusion $\Phi(\tau) \sim \frac{2}{\pi} \tau \ln \tau$ at large $\tau$. A range of superdiffusion can be obtained e.g. $c(\tau) \sim \tau^{-b} \sin(\tau+\psi)$ at large $\tau$ yields $\Phi(\tau) \sim \tau^{2-2 b}$ for $0<b<1/2$.

\begin{figure}[htpb]
  \centering
  \includegraphics[width=0.7\textwidth]{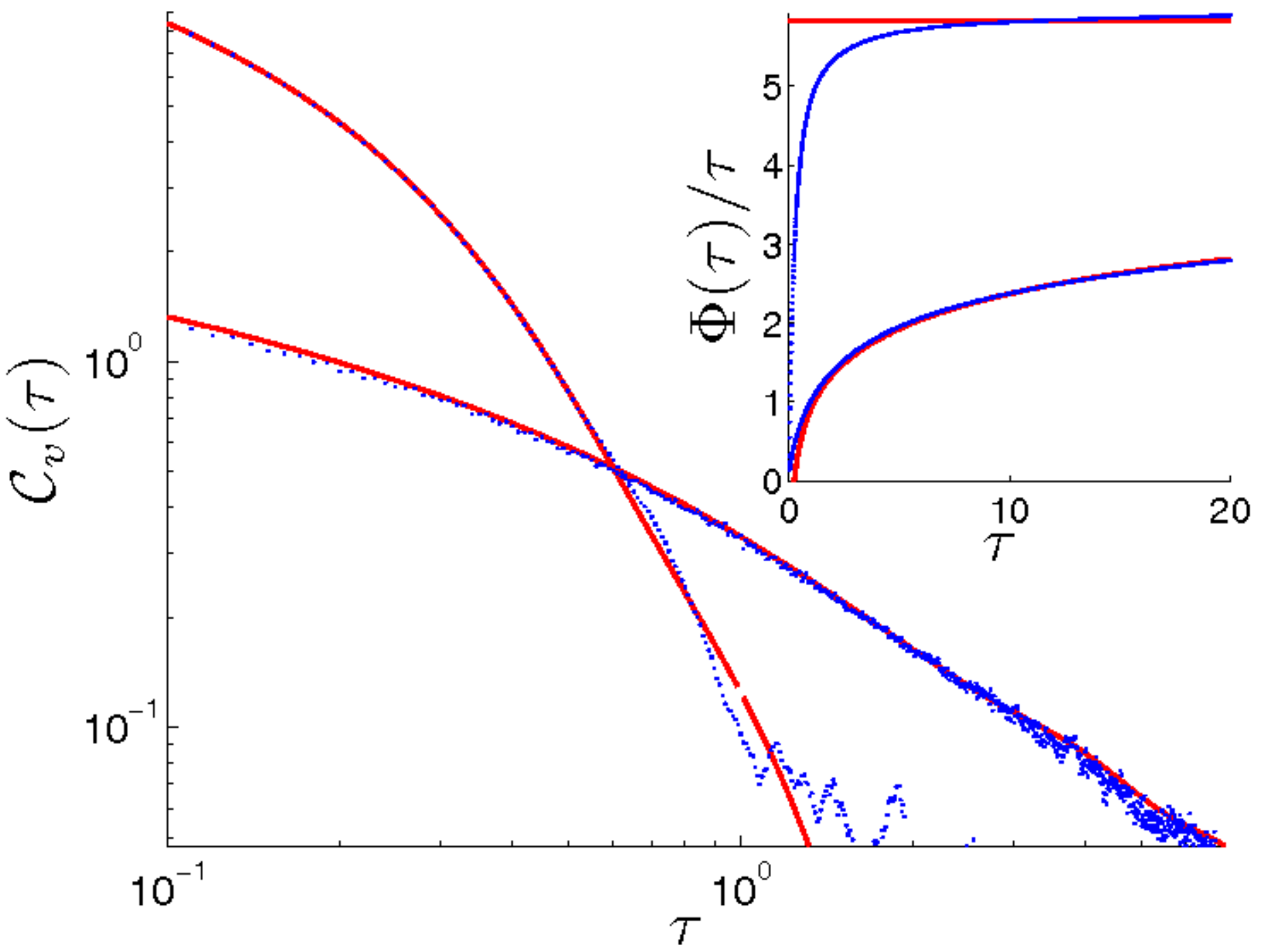}

   \caption{The angular velocity correlation function $\mathcal{C}_v(\tau)$ as a function of $\tau$ for the $c(\tau)$ used in Fig.1 (the blue curve with the stronger decay), and for  $c(\tau)=J_0(\tau)$ (the second blue curve) together with the predictions of Eq. (\ref{eqcv}) (red curves).
In the inset the winding angle variance $\Phi(\tau)$, divided by $\tau$, is displayed 
for the same two cases. From the top, the first function (in blue) saturates to its diffusive value (red line), with $D\sim 2.92$ calculated from Eq. (\ref{diffusion}). The second function (in blue) is compared with the superdiffusion prediction $\Phi(\tau)=\frac{2}{\pi} \tau \log \tau + 0.907 \tau$ from Eq.(\ref{derivative2}). Both results are an average over 50 realizations}
    \label{fig:5}
\end{figure}

\begin{figure}[h]
  \centering

  \includegraphics[width=0.42\textwidth]{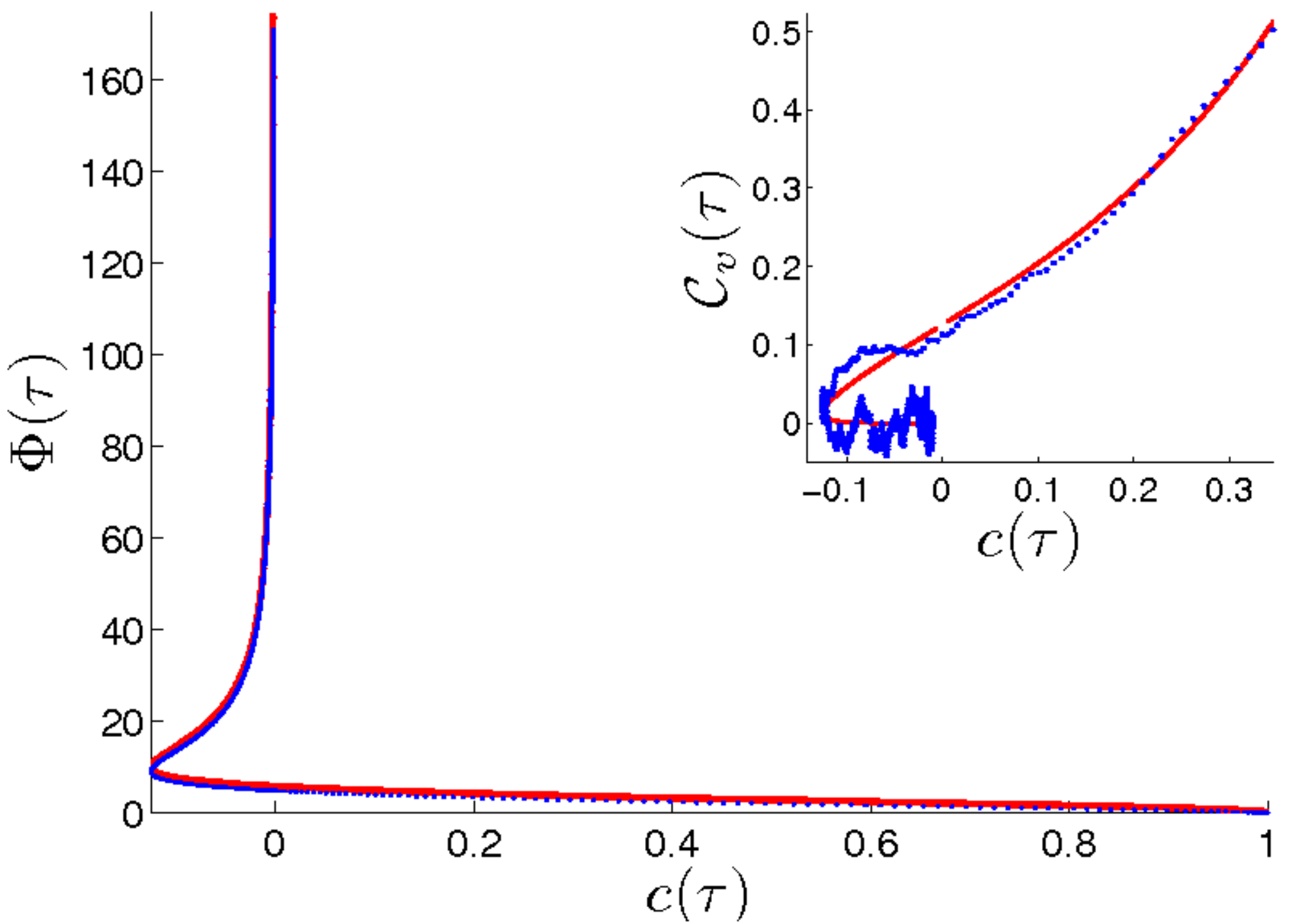}
  \includegraphics[width=0.42\textwidth]{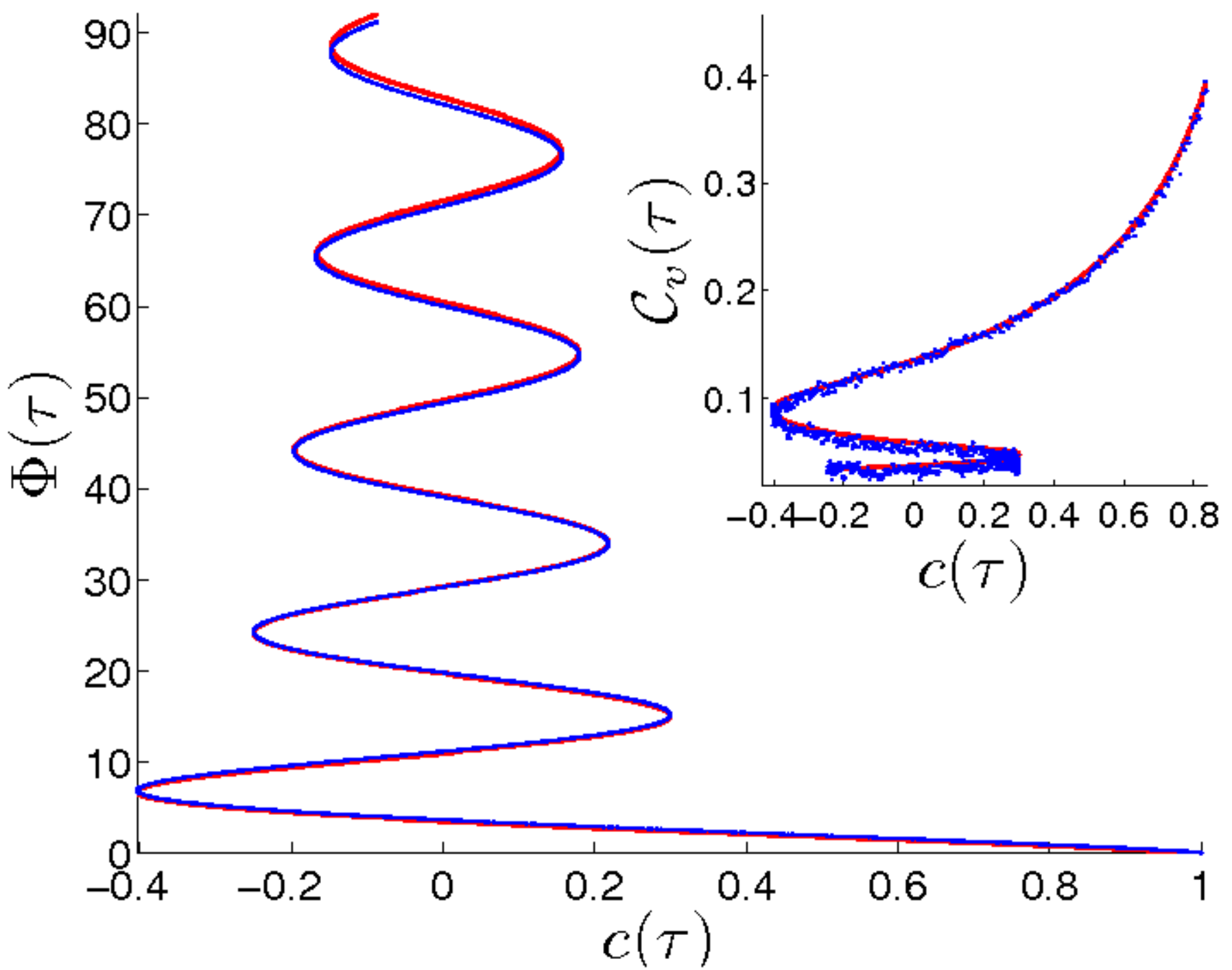}
  \caption{ {\bf Left panel: }Parametric plot of the winding angle variance $\Phi(\tau)$ ($y$-axis) and $c(\tau)$ (x-axis) for the $c(\tau)$ used in Fig.1 (blue curves) and the linear diffusion formula $\Phi(\tau)=2D\tau$ with $D\sim 2.92$ as predicted from Eq. (\ref{diffusion}) (in red). The inset shows the angular velocity correlation function $\mathcal{C}_v(\tau)$ as a function of $c(\tau)$ for for the same choice of $c(\tau)$ (blue curves) and the results of Eq. (\ref{eqcv}) (red curves).
  {\bf Right panel: } Parametric plot of the winding angle variance $\Phi(\tau)$ as a function of $c(\tau)$ for $c(\tau)=J_0(\tau)$ (blue curves) and the asymptotic
prediction $\Phi(\tau)=\frac{2}{\pi} \tau \log \tau + 0.907 \tau$ calculated from Eq. (19) (in red). In the inset the correlation function $\mathcal{C}_v(\tau)$ is shown as a function of $c(\tau)$ for the same $c(\tau)$ (blue curves), together with the results of Eq. (\ref{eqcv}) (red curves). Both results are an average over 50 realizations}
 \label{fig:3}
\end{figure}

The above predictions are checked numerically in Fig. \ref{fig:5} in the time variable $\tau$, and as a parametric plot using $c(\tau)$ in Fig. \ref{fig:3}, for the diffusive and superdiffusive case. 

Finally let us consider the stationary process $c_a(\tau)=e^{a-\sqrt{a^2 + \tau^2}}$, which as $a \to 0$ converges to the non-smooth process $C(\tau)=e^{-|\tau|}$. For any small $a>0$ the process is smooth and leads to diffusion. The diffusion coefficient however diverges as $D \approx \frac{1}{2} \ln(1/a) + \frac{\pi}{4} $ as $a \to 0$. As we will see below the non-smooth limit process $C(\tau)=e^{-|\tau|}$ is related to Brownian motion and leads to broad distributions of winding angle and to an infinite variance due to singular behaviour at short times. 

\subsection{non-stationary processes} 

We now study non-stationary processes. Such process often occur in the context of aging or coarsening dynamics \cite{aging}. In fact they can, in some cases, be mapped onto a stationary process using the property of reparametrization of time: if the process $c_{tt'}$ has a winding angle $\phi_t$ then the process $c_{g(t) g(t')}$ has a winding angle $\phi_{g(t)}$ for any positive monotonic function $g(t)$. Note that Eq. (\ref{result}) and (\ref{integrate}) have precisely the differential form required to satisfy this property. Hence for processes of the form $c_{t t'}=\hat c(g(t)-g(t'))$, the variance of the winding angle is immediately obtained as $\Phi_{tt'}=\hat \Phi(g(t)-g(t'))$ where $\hat \Phi(\tau)$ is the variance for the stationary process $\hat c(\tau)$. Hence diffusion in $\hat \Phi(\tau) \sim 2 D \tau$ implies $\Phi_{tt'} = 2 (g(t) - g(t')) D$ at well separated times. One example, frequent in aging processes, is $c_{tt'}= f(t'/t)$ for $t>t'$. Then one can choose $g(t) = \ln t$ and $\hat c(s) = f(e^{-s})$. To avoid divergences at short time difference one must have that $f(x)=1-(1-x)^b$ for $x$ close to unity, with $b>1$. One example is $f(x)=e^{-|\ln x|^b}$. If this is the case, and provided convergence at large $s$, one finds that $\Phi_{tt'} \sim 2 D \ln(t/t')$ at well separated times, i.e. diffusion in the logarithm of time. Clearly the case $b=1$ leads again to a non-smooth process and is discussed below. 

Among non-stationary processes, processes with stationary increments are of special importance. One such process is the so-called fractional Brownian motion (FBM), $C_{tt'}=\frac{1}{2}(t^{2 h}+(t')^{2 h}-|t-t'|^{2 h})$, with $0<h<1$, which is the only member of this class which is also scale invariant. For $h=1/2$ one recovers the standard Brownian motion. The FBM with $h>1/2$ is sufficiently smooth for the above considerations to apply and one easily sees that the time change $g(t)=\ln t$ and 
\begin{equation} \label{stateq} 
\hat c(s) = \cosh( h s) - 2^{2h-1} |\sinh(s/2)|^{2 h}
\end{equation}
can be used, leading to diffusion for the winding angle in the variable $g(t)=\ln t$ at large times, i.e. $\Phi_{tt'} \sim 2 D_h \ln(t/t')$ where $D_h=\int_0^\infty \hat c'(s)^2/(1-\hat c(s)^2)$ diverges as $h \to 1/2^+$. 

The cases of the Brownian motion $h=1/2$ and of the FBM for $h<1/2$ require a separate discussion. Let us first recall 
what was found for the two dimensional BM with diffusion coefficient $D_0$ i.e $\langle |\xi_t|^2 \rangle=2 C_{tt}=2 D_0 t$. At large $t$, for BM in the full plane the classical result \cite{spitzer} is that $y=2 \phi_t/ \ln t$ has a Cauchy distribution $p_0(y)=\frac{1}{\pi} (1+y^2)^{-1}$, hence the variance $\Phi_{tt'}$ cannot be defined. This broad distribution is regularized in presence of an absorbing (respectively reflecting) center in 0 of radius $R$, where the distribution of $y=2 \phi_t/ \ln (D_0 t/R^2))$ becomes $p_A(y)=\pi/(4 \cosh^2(\pi y/2)$ (resp. $p_R(y)=1/(2 \cosh(\pi y/2))$). In the two latter cases one has $\Phi_{tt'} \sim \frac{1}{12} \ln^2(D_0 t/R^2)$ at large $t$ and fixed $t'$ (resp. $\Phi_{tt'} \sim \frac{1}{4} \ln^2(D_0 t/R^2)$), however these are no more gaussian processes so the comparison with our results is not straightforward. We see below however that the process studied here yields a very similar result. Before we do so let us recall how the above scaling $\phi_t \sim \ln  t$ for the (regularized) BM results can be understood from simple arguments. From the BM properties one easily obtains the stochastic equations for radius and angle (in Ito formulation) as
$dr_t = dB_t + \frac{dt}{2 r_t}$ and $d\phi_t=d\tilde B_t/r_t$ where, for $D_0=1$, $B_t$, $\tilde B_{t}$ are two independent unit Brownian motions (i.e. with $\langle dB_t^2 \rangle=\langle d\tilde B_t^2 \rangle  = dt$). Diffusion in the winding angle is thus only possible if $\langle 1/r_t^2 \rangle$ is bounded, as also nicely discussed in \cite{comtet}. If the Brownian can explore large distance then $\langle 1/r_t^2 \rangle \sim  \ln(D_0 t/R^2)/(D_0 t)$, where the small distance cutoff is also necessary, and one recovers the above non-diffusive behaviour $\phi_t \sim \ln  t$ from the estimate $d\phi_t^2 = D_0 dt \langle 1/r_t^2 \rangle \sim 
dt \ln(D_0 t/R^2)/t$. 

Can we make contact with our results, in particular can we also obtain from our formula (\ref{result}) the (regularized) BM scaling $\Phi_{tt'} \sim \ln^2 t$ ? The answer is yes, but since we can only address smooth processes, we now consider the general smooth process with stationary increments:
\begin{eqnarray}  \label{statincr}
C_{tt'}= \frac{1}{2}( f(t) + f(t') - f(t-t'))
\end{eqnarray}
with $f''(t)=2 C_2(t)$ in the notations of the Introduction, with $f(0)=0$ hence $C_{tt}=f(t)$. The choice $f(t) \sim t$ at large $t$ corresponds to the random walk with a short time cutoff. Apart from the short times, it should look like the BM on large time scales. One example is $\xi_t=\int_0^t dt' \eta_{t'}$ where $\langle \eta_{t} \eta_{t'} \rangle =\frac{1}{2} e^{-|t-t'|}$ then $f(t)=t-1+e^{-t} \sim t^2/2$ at short times. In general $f(t)$ is an increasing function. Taking the large $t$ limit at fixed $\tau=t-t_2$ one finds:
\begin{eqnarray}  \label{contribution}
{\cal C}_v(t,t-\tau) \approx  -\frac{f''(\tau)}{2 f(t)} \ln(f(\tau)/f(t)) 
\end{eqnarray}
This gives $\partial_t \Phi_{tt'} = 2 \int_0^{t-t'} d\tau {\cal C}_v(t,t-\tau)  \approx \frac{  \ln f(t)}{f(t)} \int_0^{\infty} d\tau f''(\tau)$ hence for the random walk $f(t) \sim D_0 t$ at large $t$ one finds:
\begin{eqnarray} \label{pred}
\Phi_{tt'} \sim \frac{1}{2} (\ln t)^2 
\end{eqnarray}
and one recovers the behaviour of the regularized BM. The prefactor, however, is different from both the absorbing and reflecting core, but is nicely anticipated from the simple argument presented above. 

The same calculation can be performed for the regularized FBM for $h<1/2$, i.e. for the FBM random walk with $f(t) \sim t^{2 h}$ at large $t$ with $h<1/2$ and $f(t)$ smooth at small $t$. There one finds that the leading term above vanishes and one obtains:
\begin{eqnarray} \label{powerlaw} 
\Phi_{tt'} \sim \frac{t^{1- 2 h}}{2(1-2 h )} \int_0^\infty d\tau \frac{f'(\tau)^2}{f(\tau)}
\end{eqnarray}
i.e. a much faster growth of the variance of the winding angle.

Note that apart from the case of exact scale invariance $f(t)=t^{2 h}$, there is no time reparametrization which allows to map the problem (\ref{statincr}) with an arbitrary $f(t)$ to a stationary process. And only for $h>1/2$ the stationary process corresponding to the FBM, $f(t)=t^{2 h}$, is smooth enough so that the present results can be used: note that in that case, the contribution (\ref{contribution}) of the regime of fixed $t-t_2$ is integrable and contributes only a constant to the winding variance, while the regime $t_2/t$ fixed (usually called aging regime when dealing with two time correlations) gives the main contribution, leading to the result $\Phi_{tt'} \sim 2 D_h \ln (t/t')$ found above. Conversely, the aging regime gives exactly zero contribution for the BM $h=1/2$. Indeed then $C_{tt'}=t+t'-|t-t'|$, hence $c_{tt'}=\sqrt{t'/t}$ for $t>t'$, and $C_v(t,t')=0$ for $t>t'$ with a singularity at $t=t'$. Similarly, the aging regime is subdominant for the FBM random walk with $h \leq 1/2$. 

\begin{figure}[htpb]
  \centering
  \includegraphics[width=0.7\textwidth]{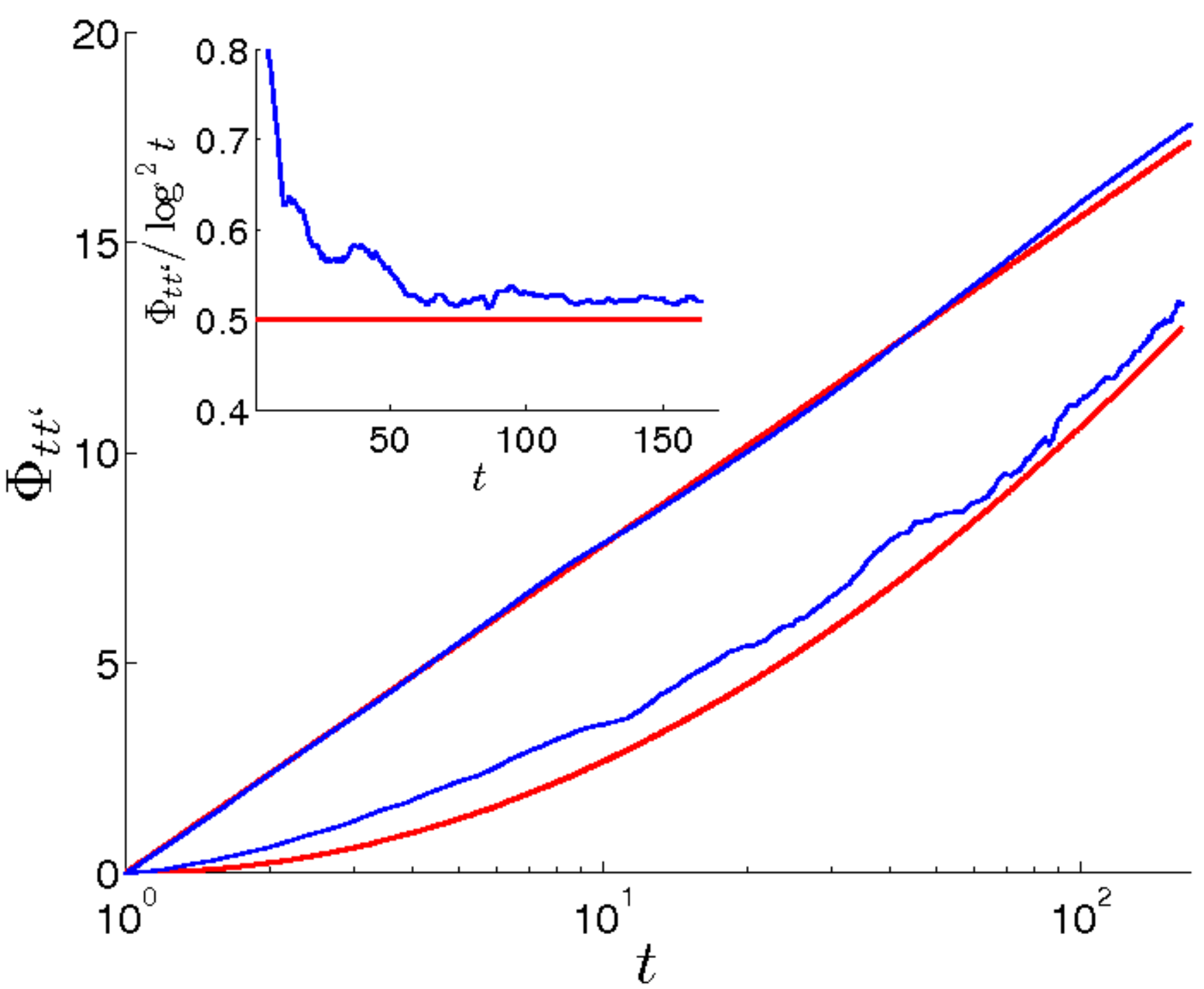}
  \caption{The variance of the winding angle $\Phi_{tt'}$ for $t'=1$ as a function of $t$ for the random walk, i.e. $C_{tt'}$ of Eq. (\ref{statincr}) with $f(t)=t-1+e^{-t}$ (bottom blue curve) as compared to the asymptotic prediction of Eq.(\ref{pred}) (corresponding red curve). The correlation functions $\Phi_{tt'}$ for $t'=1$ as a function of $t$ for the FBM with $h=0.6$  calculated using the equivalent stationary process (\ref{stateq}) with the time reparametrization $s \to e^t$ (top blue curve). The asymptotic diffusion prediction, $\Phi_{tt'}= 2 D_{h=0.6} \log t$ where Eq. (\ref{diffusion}) gives $D_{h=0.6} \approx 1.7$ is also shown (top red curve). The results for the random walk required an average over $\sim 10^3$ realizations. {\bf Inset}:The ratio $\Phi_{tt'}/\ln^2 t$ for the random walk, same data, showing the 
   convergence towards $1/2$ as predicted in (\ref{pred})}
\label{fig:6}
\end{figure}


\subsection{from Spitzer result to the winding of the stationary process $c(\tau)=e^{-|\tau|}$} 

Let us return to the unrestricted BM motion for which $C_{tt'}=\min(t,t')$ hence $c_{tt'}=\sqrt{t'/t}$ for $t>t'$. Clearly because of the singularity at $t=t'$ in $C_v(t,t')$ one cannot apply our formula for the winding angle to this case. However the formula (\ref{final}) does hold for the BM and one has for integer $n$:
\begin{eqnarray}
\langle e^{i n (\phi_t - \phi_{t'})} \rangle =F_n(\sqrt{t/t'})
\end{eqnarray}
with the large $t$ decay (at fixed $t'$) $\langle e^{i n (\phi_t - \phi_{t'})} \rangle \sim \frac{\Gamma[1+ \frac{n}{2}]^2}{n!} (t'/t)^{n/2}$. 

Using the time change $g(t)=\ln t$ one can now use the reverse correspondence and transport the Spitzer result \cite{spitzer} for planar BM to the stationary process $\hat c(\tau)=e^{-\frac{1}{2} |\tau|}$ which is not smooth and cannot be analyzed with the above methods. Hence the prediction for this process is that $y=2 \phi_{\tau}/\tau$ is distributed at large $\tau$ with the Cauchy distribution $p_0(y)$. Its variance $\Phi_{tt'}$ is thus infinite at all times, as for the BM.

An example of such a process is the Brownian motion or an ideal chain in an harmonic well, with $\tilde C(\omega)=1/(\omega^2+m^2)$, hence in real time $C(\tau)=\frac{1}{2m} e^{-m |\tau|}$ and $c(\tau)=e^{-m |\tau|}$. The distribution  of the winding for such a confined Brownian is thus again the unit Lorentzian distribution for the scaled variable $y = \phi_\tau/ m \tau$. An interesting generalization of a discrete version of this model to a chain, was studied in Ref. \cite{benichou}, with the result that, again, each monomer sees a Lorentzian winding. 

It is interesting to now consider a smoother variant of this model, i.e. an ideal chain with a small curvature energy in a harmonic well, described by $\tilde C(\omega)=(\omega^2+m^2)^{-1} - (\omega^2+M^2)^{-1} \approx 1/(\omega^4/M^2 + \omega^2+m^2)$ at large $M \gg m$. The decay in the time domain, $c(\tau)=(M e^{- m \tau} - m e^{-M \tau})/(M-m)$ is now smooth at small times, and one finds that the winding angle recovers now a finite variance and is diffusive $\Phi_{tt'} \sim 2 D \tau$ with a diffusion coefficient $D \sim \frac{1}{2} M \ln(4.2/M)$.

\section{Algebraic area enclosed} 

\begin{figure}[htpb]
  \centering
  \includegraphics[width=0.7\textwidth]{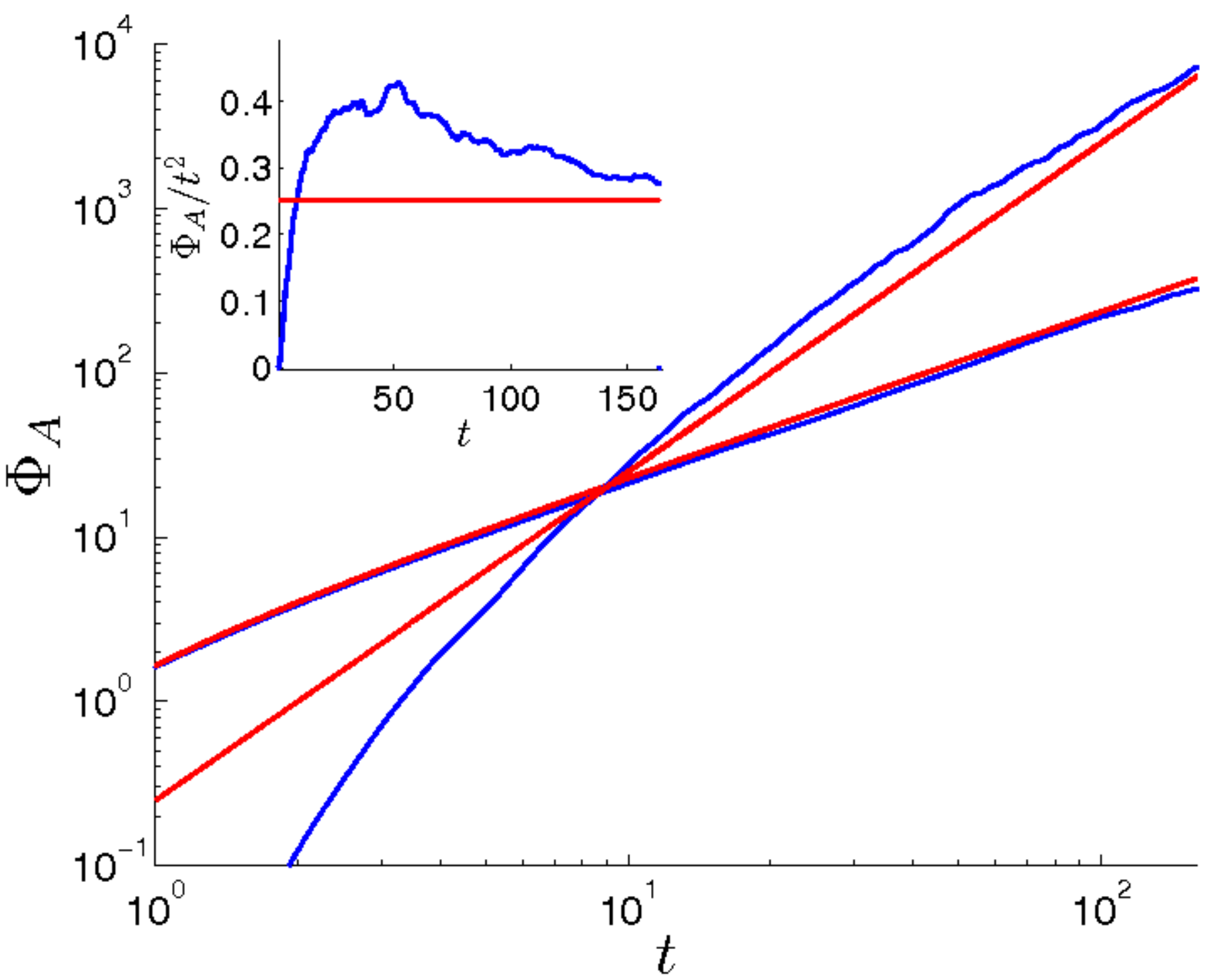}
   \caption{The variance of the algebraic area $\Phi_A=\langle \left[A_t-A_{t'}\right]^2 \rangle$ with $t'=1$ as defined above Eq.(\ref{vara}) for : (i) the random walk, i.e. $C_{tt'}$ as in Fig.5  (blue curve on top at large time) and the asymptotic prediction $\sim t^2/4$ (corresponding red curve) (ii) the result for the stationary process $c(\tau)$ of Fig.1 as a function of $t \equiv \tau$ (bottom curve) and the asymptotic prediction $2 D_A t$ with $D_A=3\pi/8$ (corresponding red curve). In the inset the ratio $\Phi_A/t^2$ is plotted for the random walk as a function of $t$, and shows saturation towards the predicted prefactor $1/4$.}
    \label{fig:7}
\end{figure}

Finally we can study the algebraic area $A_t$ enclosed by the process, which satisfies $\dot A_t=\frac{1}{2} (\xi^x_t \dot \xi^y_t - \xi^y_t \dot \xi^x_t)$. Its variance is extracted from $G_{0,0,t,t'}$ above and one easily finds that
\begin{eqnarray} \label{vara}
{\cal C}_A(t,t') = \langle \dot A_t \dot A_{t'} \rangle = \frac{1}{2} ( C^{(1,1)}_{t t'} C_{t,t'} - C^{(1,0)}_{tt'} C^{(0,1)}_{tt'} )
\end{eqnarray}
For a smooth stationary process one finds ${\cal C}_A(\tau)=-   \frac{1}{2} C''(\tau) C(\tau)+ \frac{1}{2} C'(\tau)^2$, and $\partial_\tau \langle A^2(\tau) \rangle
= 2 \int_0^\tau {\cal C}_A(\tau) =  2 \int_0^\tau C'(s)^2 ds -  C'(\tau) C(\tau)$ and one finds
the diffusion result $\langle A^2(\tau) \rangle \sim 2 D_A \tau$ with $D_A=  \int_0^\infty C'(s)^2 ds$. Let us consider now the above process with stationary increments. Note that the time reparametrization is useless here. For the random walk $f(t) \sim D_0 t$ one finds
$\langle A_t^2 \rangle \sim D_0 t^2/4$ at large $t$. This is larger than the result for Brownian paths constrained to come back to their starting points (loops) obtained in Ref. \cite{comtet}. This is well confirmed by our numerics displayed in Fig. \ref{fig:7}, where the result for a stationary process, which instead exhibits only diffusive growth of the area, is also shown for comparison.

\section{Conclusion}

We have computed here the angular velocity correlation of a very general smooth gaussian process in the plane. This allowed us to obtain a simple closed formula for the diffusion coefficient of the winding angle valid for most such stationary processes. Our formula also extends to non-stationary processes, and has allowed us to obtain the three main behaviours (i) diffusion in the logarithm of time for sufficiently smooth fractional Brownian motion (ii) the square of logarithm in time (\ref{pred}) for the winding for the usual random walks related to the Brownian motion (iii) power law growth in time (\ref{powerlaw}) for the winding angle of the random walks which provide a regularization of the non-smooth fractional Brownian motion. 

Various extensions of the present calculations are left for the future. These include: higher moments and distributions of area and winding, winding around a point different from the origin, or around several points, winding conditioned to closed paths, most general 2D gaussian process including non zero average, and finally, devising methods to account perturbatively for non gaussian effects. It would also be interesting to study persistence effects such as the probability that the winding angle never crosses zero,  or to compute the winding for loops, i.e. conditioning the process to return to its starting point. 

{\it Acknowledgments}: we thank A. Comtet for discussions, careful reading of the manuscript and pointing out Ref \cite{benichou}. This research was supported in part by the Israel Science Foundation founded by the Israel Academy of Sciences and Humanities.



\end{document}